\documentclass[letterpaper,twocolumn]{esapub} % American paper
          [1999/12/02 1.01 (PWD)]
\usepackage{times}
\usepackage{graphicx}
\usepackage{natbib}

\title{Signature of the solar cycle in the low degree p-modes using Mark-I}
\author{Sebasti\'an J. Jim\'enez-Reyes$^{1,2}$}
\author{Thierry Corbard$^{1}$}
\author{Pere L. Pall\'e$^{2}$}
\affil{$^{1}$High Altitude Observatory, NCAR, PO Box 3000, Boulder, CO 80307 USA}
\affil{$^{2}$Instituto de Astrof\'\i sica de Canarias, E-38701, La Laguna, Tenerife, Spain}

\begin{document}

%\keywords{\LaTeX; ESA; macros}

\maketitle

\begin{abstract}
High quality observations of the low degree $p$-modes exist for almost
two complete solar cycles using the solar spectrophotometer Mark-I,
located and operating at the Observatorio del Teide (Tenerife, Spain).
In this work, the observations available have been re-analyzed over a much
wider time interval than before. We analyze the time variation of the
yearly frequency shift and its frequency dependence. This information
will be used in order to average annual power spectra by removing 
the effect of the solar cycle. Using this average power spectrum, a new
estimate of the rotational splittings is attempted.
\end{abstract}

\section{Introduction}
Understanding the well known solar variability is one of the major goals
of  solar physics. In the last years, the helioseismic data,
by means of the frequency shifts of solar $p$-modes, have demonstrated 
to be very sensitive to the solar activity cycle. In particular, the existence 
of a positive shift of $\sim$0.4 $\mu$Hz, peak-to-peak,  with the solar 
cycle is very well known (R\'egulo {\it et al.} 1994, Jim\'enez-Reyes {\it et al.} 1998). 

The spectrophotomer Mark-I, has been collecting solar observations for almost 
two complete solar cycles. The available database, one of the longest
for low degrees, could be used to measure the mode parameters. However,
the variation of the central frequency is an important effect which should be taken into account in order to use long time series. 

In this work, we have analyzed the time variation of the frequency shift
and its frequency dependence. Using this information we are able to 
average the annual spectra which is then fitted in order to measure the mode
parameters, specially the rotational splittings.

\section{Analysis of the annual time series}

The data used in this work come from the observations carried out at the
Observatorio del Teide for about 15 years, from 1984 to 1999. The
observations consist of 
daily measurements of the solar radial velocity obtained with the Mark-I
resonant scattering spectrometer. Briefly, after calibration and detrending,
daily velocity residuals were joined in  consecutive 360 days leading to 
 a total of 30 time series with 
 6 months in common between consecutive series.
 The corresponding power spectra were calculated
for each one of the time series.

\begin{figure}[htbp]
\centering
\includegraphics[width=.8\linewidth]{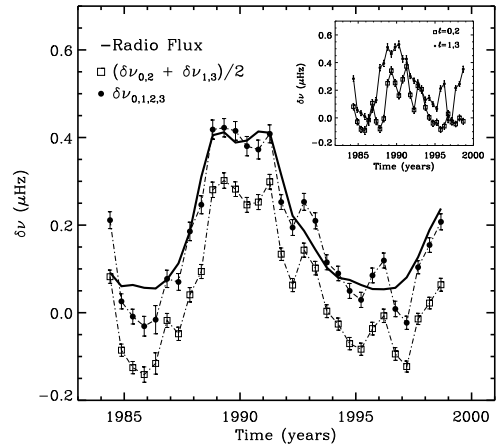}
\caption{Time variation of the frequency shift for low degree.\label{uno}}
\end{figure}

In order to determine the frequency shift, all the  power spectra have been
cross-correlated using the same reference. The position of the main
peak can be measured in different ways. 
Here, we use the fact that the cross-correlation 
between Lorentzian profiles is an other 
Lorentzian profile which  can be used to calculate the central frequency
by a least squares minimization. 

The low degree modes are asymptotically equal spaced in frequency. This 
 can be used to split in two the original spectrum, one containing
only peaks with even degrees and the other containing only peaks 
corresponding to odd degrees. Then, the same procedure as explained above was
applied to calculate the frequency shifts.

\begin{figure}[htbp]
\begin{center}
\centering
\includegraphics[width=.8\linewidth]{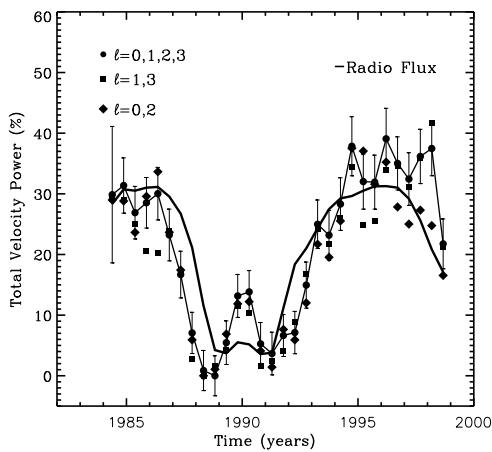}
\caption{Time variation of the total velocity power defined as the area under the main
peak of the cross-correlation function.}
\label{dos}
\end{center}
\end{figure}

In the Figure~\ref{uno} we show the integrated frequency 
shift between 2.5 and 3.7 mHz. 
The results corresponding to $\ell$=0,2 and $\ell$=1,3 have been plotted in 
the subplot at the top-right corner and its averaged (shifted -0.1$\mu$Hz) is shown in the main 
figure, which agrees very well with the integrated signal. The amplitude, 
peak-to-peak, of this changes is around 0.4$\mu$Hz, as we expected.

The best fit of the annual radio flux has also been plotted to show the good
correlation between the solar activity and the frequency shift. 
Using annual time series we have
reduced the high frequency signal 
in the variation of the frequency shift, keeping
the long term behavior. Thus, the correlation levels between these parameters
are clearly close to 1.

We have also calculated the total velocity power 
defined here as the area under the
main peak of the cross-correlation function. Once the main peak is fitted
as a Lorentzian profile, the area can be calculated from the amplitude and
linewidth.

The first demonstration of the variation for the velocity power 
for all measured  
$p$-modes with the solar activity was reported by Pall\'e {\it et al.} (1990) where 
an increase of 30 to 40 $\%$, anti-correlated with the solar 
activity cycle, was found. Afterwards, Anguera {\it et al.} (1992) found similar
results following different techniques. Those results were interpreted 
in terms of a change in the efficiency of the excitation of such modes. 
Other possibility should be the absorption of mode power by magnetic structures
like sun spots, active regions, etc. However they represent
a small influence and it cannot explain totally such high ratio.

In the Figure~\ref{dos}, we present the time variation of the 
total velocity power. Again,
we show the radio flux scaled which denotes here the behavior
of the solar activity cycle. From the figure, we are able to see the
variation between minimum and maximum of the solar cycle
is around 35$\%$ which agrees with previous results. Moreover, it
is clear the high anti-correlation with the solar cycle, thus the 
velocity power
decreases when the activity increase. The corresponding variation for
the odd and even degrees are very similar and they are shown as well in
the same figure.

\begin{figure}[htbp]
\centering
\includegraphics[width=.9\linewidth]{grad_shift_hn.epsi}
\caption{Variation of the dependence of the frequency shift like a
function of the frequency. \label{tres}}
\end{figure}

Before completing this section, we analyze the frequency dependence of the 
frequency shift. Thus, we split each one of the spectra in regions of
135$\mu$Hz which contain a set of modes $\ell$=0,1,2 and 3. Then, every 
region is cross-correlated with the same reference and finally
the method explained above is used to calculate the 
frequency shift. The results can be seen in the Figure~\ref{tres} which have been
plotted together with the best fit of the inverse mode mass averaged for
low degree and over the same region in frequency. 
As we can see, the frequency shift can be reproduced very well by the
inverse mode mass, which means that 
the origin of this variation should be close
to the solar surface.  This variation in the frequency dependence
agrees very well with earlier studies carried out at higher degree
(Libbrecht and Woodard, 1990)

The results are in good agreement with the individual fittings carried out
by Jim\'enez-Reyes (2000) shown as well in the figure.
 He has applied similar techniques at high
frequency as well, where a strong fluctuation of the frequency shift appears,
as we can see in Figure~\ref{tres} (bottom).

\section{The rotational splittings}
One way to increase the signal to noise ratio is to average
 spectra. Nevertheless, it
cannot be done directly because the central frequency of the acoustic modes
are expected to change. In that case, the linewidth would be bigger and 
the structure of the peaks are not well defined leading to wrong
fits.

\begin{figure}[htbp]
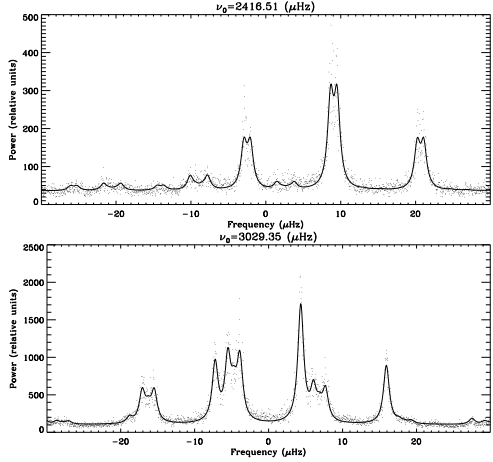

\centering
\includegraphics[width=.8\linewidth]{2416.epsi}
\includegraphics[width=.8\linewidth]{3029.epsi}
\caption{Examples of two sections of the averaged spectra. In the top a 
piece of the spectra containing a couple of odd modes $\ell$=1,3 is
successfully fitted (solid line). In the lower figure, other piece of the 
averaged spectra is showed, in this occasion for $\ell$=0,2.\label{cuatro}}
\end{figure}

Thus, before averaging spectra, we need to shift them in order to
remove the effect of the solar activity. We studied in the last
section that the central frequency of the acoustic modes changes
are a function of two parameters: time and frequency. The time dependence 
can be expressed as a function of the radio flux for instance, 
thanks to the good
correlation between them. The behavior with
the frequency is expected to be very soft, being null at low frequency
and increasing slowly at high frequency. We have also shown that
this dependence can be expressed as a function of the inverse of the
mode mass. In brief, our 30 spectra covering 15 years of the solar cycle
have been shifted and then averaged using the information obtained 
about the behavior of the frequency shift.

The Figure~\ref{cuatro} shows two different pieces of the averaged spectrum,
once each one of the yearly spectra has been shifted to remove
the effect of the solar activity. In the top, a couple of odd modes and
the best fit of our model is plotted while in the bottom a group of even
modes and its best fit is shown. Note, that the fine structures are quite
easy to identify. So, the rotational splittings is in all the cases very clear
allowing us to fit the mode parameters easily. The
sidebands emerge also from the noise and it was necessary to consider them
as well in the model. 

The statistic of the power spectra is expected to
be $\chi^{2}$ with 2 degree of freedom. However,
when many spectra are averaged the statistic becomes Gaussian, especially 
when the number of spectra averaged is large like in our case. Thus, we use 
a Marquard-Levenberg method for this work to perform a non-linear least
squares minimization, and the formal errors bars are the output of the
program.

\begin{figure}[htbp]
\centering
\includegraphics[width=.6\linewidth,angle=90]{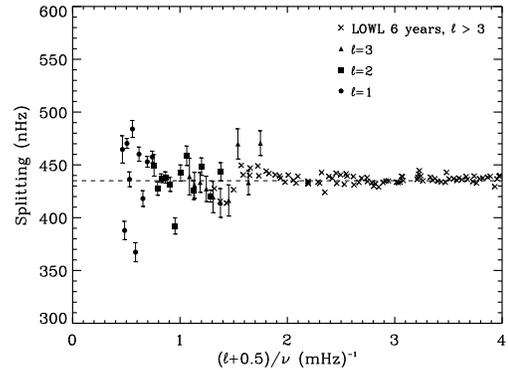}
\caption{Rotational splittings for low degree ($\ell$$\leq$3). 
The crossed represent the averaged $a_1$ coefficient using 6 years of LOWL
data. The dashed line shows a constant rotation rate at 435nHz.\label{quinto}}
\end{figure}

The advantage of the technique proposed here is in the fact the signal
is cleaned from the re-excitation component and thus become  less
spiky. In addition, structure from the
acoustic modes emerges from the noise allowing a better identification of 
the signal, which is specially important for $\ell$=3 due to the low 
sensitivity of the instrument to that degree. However, for the same reason
the second sideband due to modulation of one day, appear clearly 
in the spectra.

Adjacent degrees have been fitted together using a window 60$\mu$Hz large,
centered at halfway between the central frequency of both modes. 
The linewidth and the level of background is supposed
to be the same. It is necessary due to the distance of the these modes,
which decrease at higher frequency. This difficulty becomes more critical
in our case due to the presence of sidebands. 
\begin{table}
\begin{center}
\caption{Low degree p-mode splittings in nHz.}
\begin{tabular}{lccc} 
\hline
$n$  & $\ell$=1 & $\ell$=2 & $\ell$=3 \\
\hline
11 &  			& 443.47$\pm$ 8.60 & \\
12 &                    & 420.51$\pm$ 5.75 & 470.57$\pm$11.74 \\
13 & 457.48$\pm$ 5.51 & 448.36$\pm$ 8.23 & 433.10$\pm$11.36 \\
14 & 452.92$\pm$ 5.10 & 426.11$\pm$ 8.23 & 469.85$\pm$14.29 \\
15 & 418.08$\pm$ 7.45 & 458.62$\pm$ 9.19 & 416.30$\pm$14.76 \\
16 & 460.14$\pm$ 6.73 & 442.28$\pm$ 7.65 & 413.91$\pm$13.63 \\
17 & 367.32$\pm$ 9.00 & 392.20$\pm$ 7.69 & 419.56$\pm$14.83 \\
18 & 484.02$\pm$ 8.06 & 431.51$\pm$ 6.64 & 427.28$\pm$12.00 \\
19 & 436.31$\pm$ 7.02 & 438.07$\pm$ 5.33 & 433.35$\pm$ 9.37 \\
20 & 470.37$\pm$ 4.64 & 437.41$\pm$ 4.03 & 430.67$\pm$12.26 \\
21 & 387.85$\pm$ 8.69 & 427.49$\pm$ 6.26 & 438.90$\pm$17.38 \\
22 & 464.63$\pm$12.93 & 449.52$\pm$10.16 & \\
\hline
\hline
& 447.54$\pm$2.11 & 433.41$\pm$1.92 &  435.90$\pm$3.99\\
\hline
\hline
\end{tabular}
\label{table_splittings_int}
\end{center}
\end{table}

Finally, we present in Figure~\ref{quinto}
the rotational splittings calculated in this work
plus the average of 6 years using LOWL data (Jim\'enez-Reyes 2000), for 
degree above four. A straight line
at 435 nHz has been drawn over the data. The rotation rate
seems to be constant, however at higher depth, 
there is a certain scattering but no clear  trend 
can be seen. The successful fits are also tabulated in the Table 1. 
The last row represents the weighted mean for each degree.
All these results lead us to think that the rotation
rate in the core is very likely to be 
constant and equal to the rest of the radiative
part, although some dispersion remain at very low turning point.

We have compared our results which those obtained by Bertello {\it et al.} (2000)
and Chaplin {\it et al.} (1999). The first work describe the analysis carried
out on a time serie from 759 days of calibrated disk-averaged velocity
signal provided by two different instruments, GOLF and MDI, both on board
the Solar and Heliospheric Observatory (SOHO). 
This data begin close to the solar minimum in 1996 May 25,
where any important change in the central frequency of the acoustic
modes is expected, and finish in 1998 June 22. 
Moreover, they have considered two different kind of
fits, one using the typical Lorentzian profile and other which introduce
possible asymmetries.

\begin{figure}[htbp]
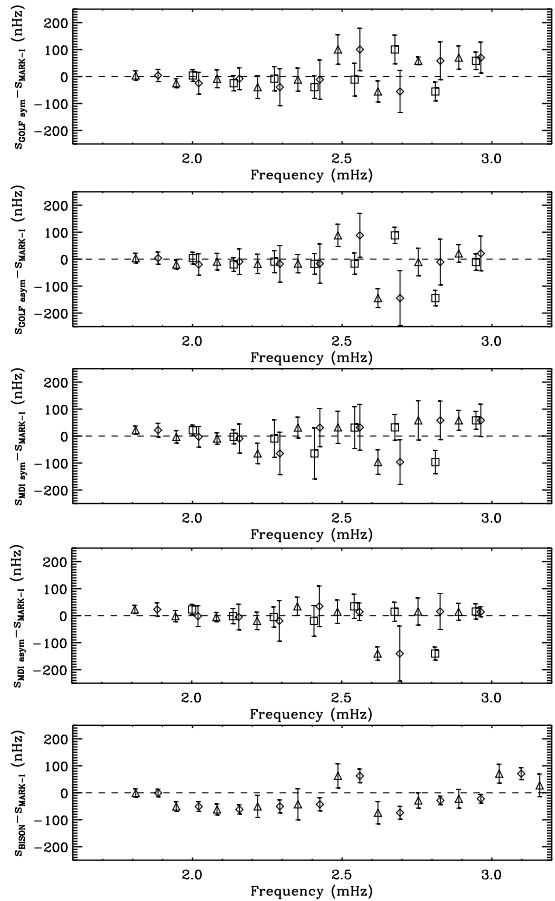

\centering
\includegraphics[width=.9\linewidth]{difsplit1.epsi}
\includegraphics[width=.9\linewidth]{difsplit2.epsi}
\includegraphics[width=.9\linewidth]{difsplit3.epsi}
\includegraphics[width=.9\linewidth]{difsplit4.epsi}
\includegraphics[width=.9\linewidth]{difsplit5.epsi}
\caption{Differences of the rotational splittings in nHz between others works and those
measured in the present analysis.\label{sexto}}
\end{figure}

The second work represents the last published splittings measured for
BISON team. They come from of a 32-month power spectrum 
generated from velocity signal collected between 1994 May 16 and 1997
January 10. Again, the period considered is near the solar activity
minimum. In both works the duty cycle is higher than in our case, being
close to 100$\%$ for GOLF and MDI.

The differences of our work with those works are shown 
in the Figure~\ref{sexto}. 
%There is a 
%systematic difference when the asymmetric profile is considered. Thus,
%the differents are slightly positive in that case and slightly negative
%when a symmetric profile is considered. 
%5In any case, there is a good
%agreement and only and high frequency the differents are more 
%significative.
There are significant differences 
between our results and those obtained by BISON group. Their
splittings are in general lower than our results. However, this 
does not appear in the comparison with  GOLF or MDI data,
 for which the differences
are more random around zero with a small scattering, especially 
at low frequency. These differences are specially small when
a asymmetric profile were considered.

\section{Conclusion}
An analysis of the $p$-mode frequency shift over more than one solar
cycle has been carried out. The time dependence of the frequency
shift as well as its frequency dependence have been parameterized.
Then, this information has been used in order to shift each one of the annual
spectra, removing therefore the signature of the solar activity cycle.

The rotational splittings are in good agreement with those obtained
by Bertello {\it et al.} (2000). Thus, the rotation rate is very close to
435 nHz like the rest of the radiative part. However, there is still
some fluctuation at lower turning point. The comparison with the 
splittings measured by Chaplin {\it et al.} (1999) shows a significant
difference.

The method presented here can be useful in the analysis of long time
series, covering an important part of the solar cycle. Last but not
least, if all the ground instruments (e.g. BiSON, IRIS, ECHO) are joined,
then the quality and also the duty cycle of the time series will increase,
reducing the systematic errors, allowing new perspectives in the
analysis of the solar activity cycle using the frequency shift as
a new solar index.

\section{Acknowledgments}
We are extremely thankful to Mausumi Dikpati for useful discussions
and additional comments. S.J. Jim\'enez-Reyes and T. Corbard are very 
thankful to the organizers of the meeting for providing financial support.
T. Corbard acknowledge support from NASA grant S-92678-F.

% The following bibliography was produced with
%   \bibliographystyle{aa}
%   \bibliography{esapub}
% The results are inserted directly here to simplify
% the demonstration.

\end{document}